\documentclass[aps,prl,twocolumn,showpacs]{revtex4}
\usepackage{graphicx}
\usepackage{dcolumn}
\usepackage{bm}

\begin{document}


\title{Role of commensurate arrangements in the optical response of metallic gratings.}


\author{A. Barbara, J. Le Perchec, P. Qu\'{e}merais, T. L\'{o}pez-R\'{i}os}
\email[]{aude.barbara@grenoble.cnrs.fr}
\altaffiliation{}
 \affiliation{Institut N\'{e}el (CNRS/UJF), 25 avenue des Martyrs, BP 166, 38042 Grenoble Cedex 9, France}
\author{S. Collin, C. Sauvan, J-L. Pelouard}
 \affiliation{Laboratoire de Photonique et Nanostructures (LPN/CNRS), Route de Nozay, 91460 Marcoussis, France}

\date{\today}

\begin{abstract}
Light localization on commensurate arrangements of metallic
sub-wavelength grooves is studied. We theoretically show that as the
degree of commensuration tends to an irrational number new light
localization states are produced. These have properties close to
that reported for hot spots on disordered surfaces and are not
permitted for simple period gratings. Some theoretical predictions
are experimentally provided in the infra-red region by reflectivity
measurements performed on two commensurate samples with respectively
two and three slits per period.
\end{abstract}


\maketitle Metallic systems such as random surfaces, thin films
deposited on cold substrate or gratings, are known for their various
and sometimes puzzling optical properties. In another context it is
also known that quasi-crystals giving rise to a "forbidden"
five-fold symmetry as well as one dimensional incommensurate
structures may present electron localization
effects\cite{mayou1,mayou2,mayou3,aubry}. We thus expect new light
localization effect by merging electromagnetic resonances of
metallic surfaces and 1D incommensurate structures. The arrangements
we study are sketched on fig.1. They are composed of several
identical closed rectangular grooves per period separated by two
possible sub-wavelength distances $L$ (for long distance) and $S$
(for short distance). The sequence of lengths, $LSSLS$ for instance,
is chosen to be uniform and the arrangement is
commensurate\cite{aubry,ducastelle,quemerais}. We can by this way
study the optical properties of surfaces with complex topologies and
interestingly we find that they present strong and sharp resonances
associated to very local near-field intensity enhancements. Systems
with $N$ grooves per period, but separated by the same distance,
have been theoretically studied\cite{skigin,skigin2}. In the present
context, they can be seen as particular cases of our commensurate
structures given by the sequence of lengths $SSS..(\times N)L$. Up
to three cavities per period, both approaches are equivalent.
However, as grooves continue to be added our arrangements tend to a
quasi-periodic (incommensurate) structure and both the near and far
field properties become very different. Indeed, the commensurate
arrangements present local, strong and wavelength dependency of the
field localization inside the grooves. The near-field intensity
distributions are associated to sharp cavity resonances which can be
evidenced by dips in the reflectivity. An experimental prove of
their existence is presented via reflectivity measurements performed
in the infra-red region on gratings with respectively two and three
slits per period. In that particular case, these measured modes also
give the first experimental evidence of those predicted by Skigin et
al.\cite{skigin3}.
\begin{figure}
\centering\includegraphics[width=7.5cm,height=5cm]{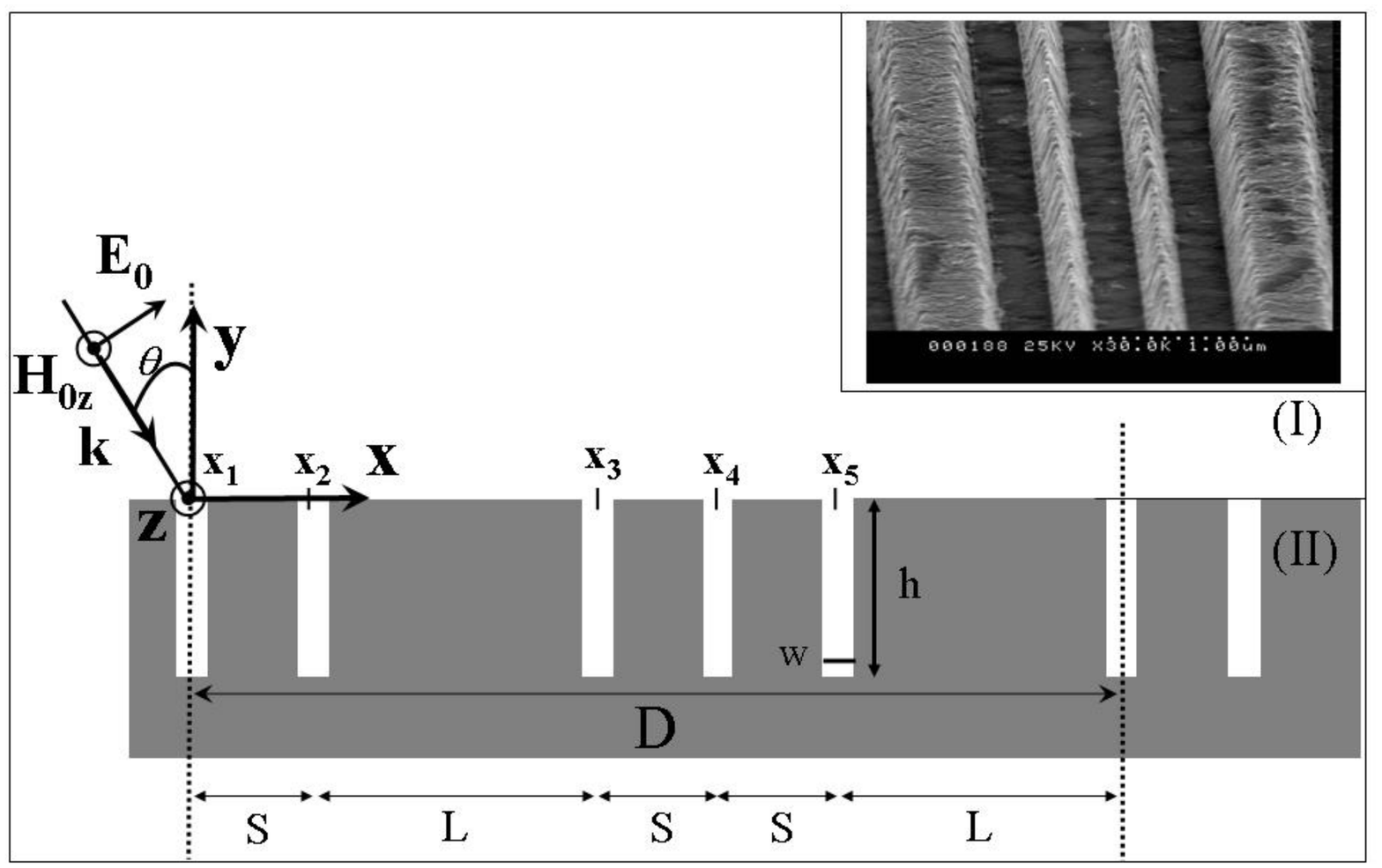}
\caption{Example of a uniformly ordered commensurate structure
$<3/5>$ of period $D=3S+2L$ where $L$ and $S$ respectively stand for
the long and short distances separating two cavities. The grooves
are identical with an aperture $w$ and a height $h$. Inset
represents a SEM image of the $<2/3>$ sample.}
\end{figure}\\The gratings of interest are characterized by a series
$\{x_n\}_{1\leq n\leq Q}$ of location of the center of the $Q$
cavities contained in a period. The structures are generated fixing
the origin at $x_1=0$ and using the equation\cite{quemerais}:
\begin{eqnarray}
&x_{p+1}&=x_p+S\times\sigma_p+L\times(1-\sigma_p)\nonumber
\\ \text{with} &\sigma_p&=[p\times R/Q]-[(p-1)\times R/Q] \nonumber
\end{eqnarray}
where the function $[y]$ defines the integer part of $y$, $R$ the
number of grooves separated by the distance $S$ and $Q$ the number
of grooves per period. These commensurate structures are called
$R/Q$ gratings, their period is $D=R \times S + (Q-R) \times L$ and
the ratio $R/Q$ indicates the order of commensurability. It is
chosen to be an irreducible rational number. For example, the
successive $R/Q$ may be built by using the Fibonacci series
$\{1,2,3,5,8,13...\}$ and take de values $\{1/2,2/3,3/5,5/8...\}$.
For large $R$ and $Q$ the structure tends to a quasi-periodic (or
incommensurate) one\cite{quemerais}, $D\rightarrow\infty$, and
$R/Q\rightarrow\xi=(\sqrt{5}-1)/2$ the inverse of the mean gold
number. We have calculated the optical response of these structures
illuminated by a p-polarized plane wave (magnetic field parallel to
the groove) following the procedure detailed in ref\cite{nousEPJD}.
The theoretical approach is an approximated modal method using the
surface impedance boundary conditions\cite{wirgin}, which has been
several times successfully compared with experiments\cite{jap} when
considering gold gratings with sub-micron geometrical parameters in
the infra-red region. For commensurate gratings of period $D$ the
Rayleigh development of the field in the region (I) above the
grating (fig.1) is:
\begin{displaymath}
H_{z}^{(I)}(x,y) ={e^{ik(\gamma_0 x-\beta_0
y)}+\sum^{m=+\infty}_{m=-\infty}R_m e^{ik(\gamma_mx+\beta_my)}},
\end{displaymath}
where $k=2\pi/\lambda$ is the wave vector of the incident plane
wave. The terms $\gamma_m=sin\theta+m\lambda/D$, $\beta_m$ defined
as $\beta_m^2=1-\gamma_m^2$, and $R_m$ are respectively the
normalized wave vectors and the amplitude of the $m^{th}$ order of
reflection. In region (II) the field is determined considering the
vertical walls as perfectly conducting and applying the surface
impedance approximation at the bottom of the cavity:
\begin{eqnarray}
H_{z}^{(II)}(x,y)&=&\sum_{p=1}^Q\sum_{n=0}^{+\infty}A_p^n \cos \left[\frac{n\pi}{w}\left(x-x_p+\frac{w}{2} \right )\right]\nonumber \\
&\times& {\left( e^{i \mu_n (y+2h)}+ r_n e^{-i \mu_n y}
\right)\sqcap(x-x_p)}
\end{eqnarray}
where $\sqcap(x)=1$ for $x\in[-w/2,w/2]$ and $0$ elsewhere. $A_p^n$
is the amplitude of the $n^{th}$ mode in the $p^{th}$ cavity with
$\mu_n=k\sqrt{1-(\frac{n\lambda}{2w})^2}$. The terms
$r_n=(\mu_n/k+Z)/(\mu_n/k-Z)$, where $Z=1/\sqrt\varepsilon$ and
$\varepsilon$ the dielectric constant of the metal, are the
reflection coefficients of the $n^{th}$ mode at the bottom of the
cavity. Determining the coefficients $A_p^n$ and $R_n$ we find the
electromagnetic field in the whole space. $A_p^n$ is obtained by
solving the matrix system:
\begin{displaymath}
\label{eqMatrix} \sum_{\ell=0}^{+\infty}\sum_{p'=1}^Q
M_{n,\ell,p,p'}A^\ell_{p'}=V_p^n
\end{displaymath}
\begin{eqnarray}
\text{with} \qquad
{V_p^n}&=&{\left(\frac{2}{1+\delta_{n,0}}\right)\frac{2\beta_0}{\beta_0+Z}S^+_{0n}e^{ik\gamma_0x_p}}
\nonumber
\\ \nonumber
\\ \nonumber
 \text{and} \qquad {M_{n,\ell,p,p'}}&=&{(e^{2i\mu_nh}+r_n)\delta_{n,\ell}\delta_{p,p'}}
\nonumber
\\
&-&{F_{n\ell}\sum_{m=-\infty}^{+\infty}\frac{S_{mn}^+S_{m\ell}^-}{\beta_m+Z}
 e^{ik\gamma_m(x_p-x_{p'})} \nonumber}
\end{eqnarray}

where we have defined the terms $S_{mn}^{\pm}$ and $F_{n\ell}$ as:
\begin{eqnarray}
{S_{mn}^\pm}&=&{\frac{1}{w}\int_{-w/2}^{+w/2}e^{\pm ik\gamma_nx}\cos\left[\frac{m\pi}{w}\left(x+\frac{w}{2}\right) \right]dx}, \nonumber \\
{F_{n\ell}}&=&{\left(\frac{2}{1+\delta_{n,0}}\right)\frac{w}{D}(e^{2i\mu_{\ell}h}-1)\left(\frac{\mu_{\ell}}{k}+Z\right)};\nonumber
\end{eqnarray}

while $R_n$ is given by:
\begin{eqnarray}
{R_m}&=&{\left(\frac{\beta_0-Z}{\beta_0+Z}\right)
\delta_{m,0}+\frac{w}{D}\sum_{p=1}^Q\sum_{n=0}^{+\infty}A_p^n
e^{-ik\gamma_nx_p}\times} \nonumber \\
&\times&
{S_{mn}^-\left(\frac{\mu_n/k+Z}{\beta_m+Z}\right)(e^{2i\mu_nh}-1).}\nonumber
\end{eqnarray}
\begin{figure}
\centering\includegraphics[width=8.5cm,height=8.5cm]{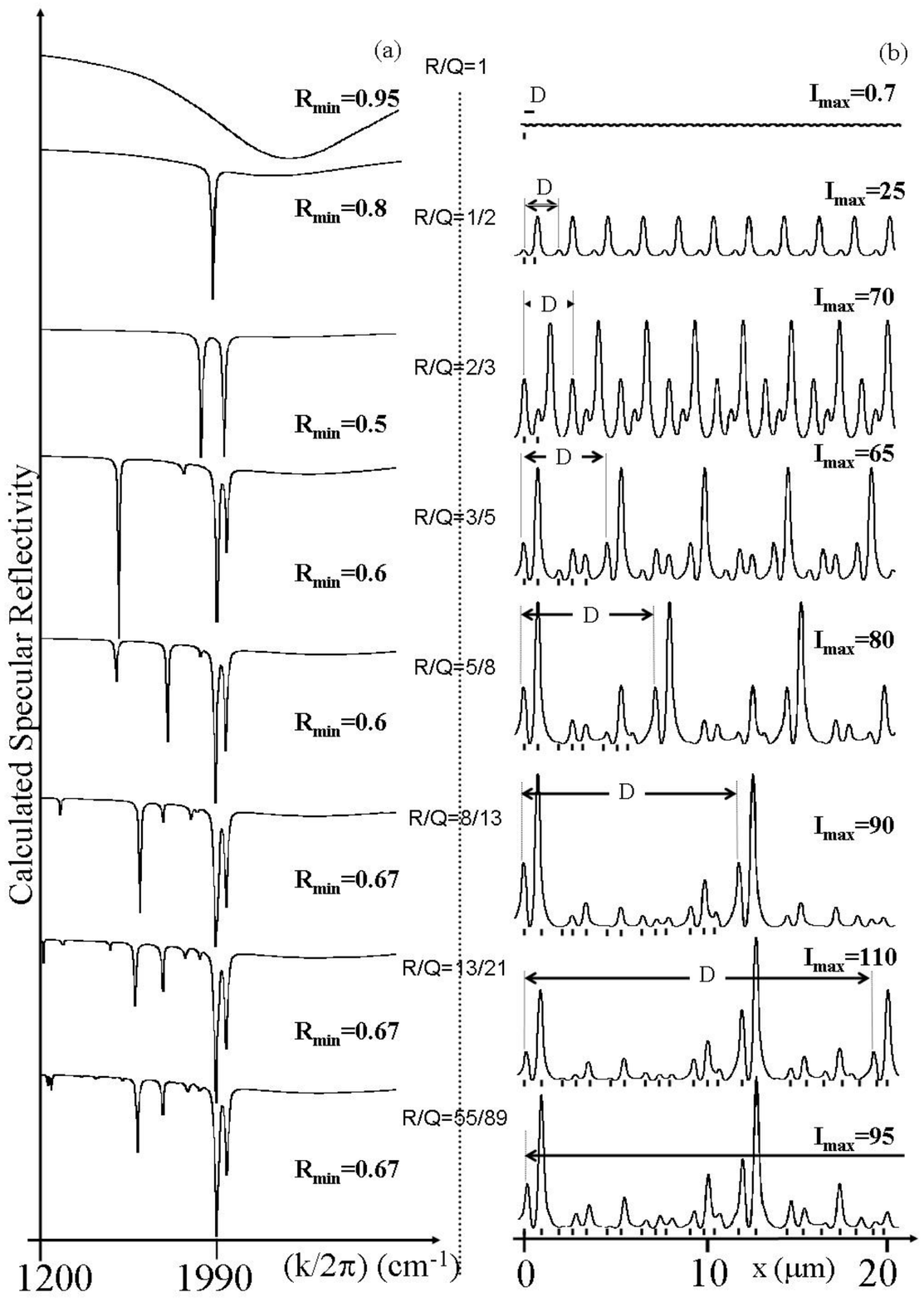}
\caption{(a) Specular reflectivity and (b) magnetic field intensity
at the interface $y=0$ at the frequency resonance located around
$1990$ $cm^{-1}$, calculated for several uniform structures as $R/Q$
tends to the inverse of the mean golden number. Simulations were
made with $h=1 \mu m$, $w=0.3 \mu m$, $S=0.7 \mu m$, $L=1.2 \mu m$
and an incidence angle $\theta=15^\circ$. Vertical black lines in
(b) indicate the position of the cavities within one period.}
\label{fig-gratings}
\end{figure}
Figure 2a displays the evolution of the specular reflectivity
calculated at $\theta=15^\circ$ as the order of commensurability
tends to $\xi$. Simulations were made considering only the
fundamental mode (n=0 in eq.1) and using the gold dielectric
constant from ref\cite{palik}. For a simple-period grating, the
spectrum presents the well-known broad dip due to the excitation of
the Fabry-Perot like resonance inside the cavities\cite{lopez}. As
grooves are added, by considering the successive commensurate
structures ${1/2,2/3,3/5,5/8}$, supplementary and sharp resonances
appear. For larger Q ($Q>8$) only slight changes appear in the
spectrum; the optical properties converge. In particular, one can
see how the $<8/13>$ and the $<55/89>$ arrangements almost exhibit
the same features. This is a direct indication of the
self-similarity of the structures produced as $R/Q\rightarrow
\xi$\cite{ducastelle}. The convergency of the optical properties
lead to a situation where some strong and narrow resonances persist
and are associated to strong local fields. These properties are also
direct consequences of the (in)commensurate character of the
structure and do not exist for simple or compound gratings. Fig.2b
shows the progressive establishment of hot spots, which are features
experimentally observed on disordered metallic
surfaces\cite{gresillon}, as $R/Q\rightarrow \xi$. In the chosen
configuration, the magnetic field intensity at the interface can
locally be more than two orders of magnitude larger than the
incident one (fig.2b) and this corresponds to enhancements of the
electric field at the interface, larger than three orders of
magnitude.
\\ The occurrence of these new sharp cavity modes is due to the breaking
of symmetry of the system. The pseudo-periodicity of the fields
remains valid for the global period $D$ such that cavities within a
period do not see the same exciting field and this opens up for
different configurations. A physical image can be given by analogy
to coupled dipoles. Indeed, in the case of a two-groove system a
complete analytical study of the fields expression can be performed
and allows to quantitatively characterize the nature of the
resonances\cite{nousPRL}: each resonating cavity acts as a damped
oscillating dipole and their near-field coupling causes the
splitting of each individual mode into a large one corresponding to
the symmetric coupling of the cavities ($\rightarrow\rightarrow$)
and a thin one, corresponding to the anti-symmetric coupling
($\rightarrow\leftarrow$). In the same manner, $R/Q$ arrangements
act as Q oscillating coupled dipoles, each mode corresponding to a
specific configuration of the dipoles' orientation leading to more
or less radiative modes and more or less intense near-field
enhancements. Coupled modes of three slits with different individual
frequency resonances were measured in the microwave
regime\cite{sambles}. Here, the $Q$ cavities are identical but can
still resonate at slightly different frequencies. A consequence is
that the local intensity enhancements can be quite critically
wavelength dependent. This property is also experimentally observed
on disordered\cite{gresillon} or SERS (Surface Enhanced Raman
Scattering) surface\cite{moskovitz}. This point is illustrated in
fig.3: maps of the electric field intensity along the x-axis and
above a unit cell of a $<8/13>$ grating are calculated for different
incident wave-number scanned around the strong resonance located at
1990 $cm^{-1}$. Within a very narrow spectral range, here 1995 to
2038 $cm^{-1}$, the spatial localization of the electromagnetic
field is strongly modified. For specific frequencies, one or few
cavity(ies) act as active sites by almost individually concentrating
very strong field intensity while the neighbouring ones are
extinguished (fig3a and 3c).
\begin{figure}
\includegraphics[width=8.5cm,height=5.5cm]{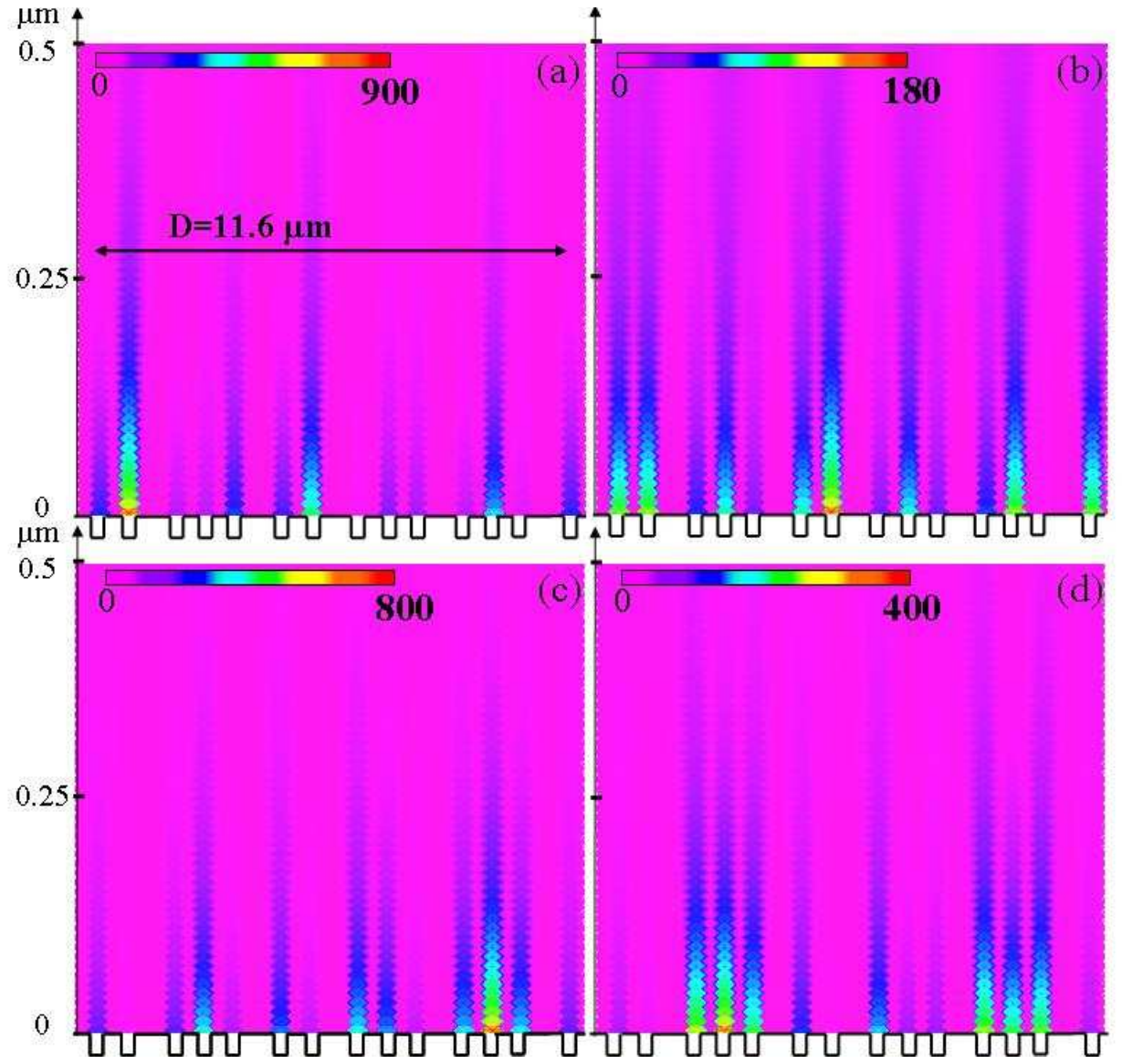} \caption{Intensity maps of the electric field along the x-axis above one period of a $<8/13>$
grating calculated at 1995 $cm^{-1}$(a) 2013 $cm^{-1}$(b) 2031
$cm^{-1}$(c) and 2038 $cm^{-1}$(d), with $\theta=10^\circ$. $S=0.7
\mu m$ and $L=1.2 \mu m$ $h=1 \mu m$ and $w=0.3\mu m$. The "jump" of
photons localization is clearly seen, while the incident wave number
varies of only a few percent.}
\end{figure}
\\
To validate some of these theoretical results, we evidenced the
existence of the sharp modes by measuring the specular reflectivity
of gratings with one, two and three slits per period. The samples
were prepared by electron-beam lithography and a double lift-off
technique similar to ref.\cite{collin}: A Si substrate is first
covered by a $200$ nm-thick gold layer and secondly by a 1
$\rm{\mu}m$-thick $\rm{Si}_3\rm{N}_4$ layer deposited by
plasma-enhanced chemical vapor deposition. $\rm{Si}_3\rm{N}_4$ walls
of the width of the slits and separated either by the distance L=1.2
or 1.5$\rm{\mu m}$ or S=0.7$\rm{\mu m}$, are performed by
electron-beam lithography and reactive ionic etching. A $650$
nm-thick gold layer is then deposited, and the $\rm{Si}_3\rm{N}_4$
walls are lifted-off by means of a HF solution. The obtained grooves
have a trapezoidal shape ($w_{down}=0.36 \mu$m and $w_{top}=0.62
\mu$m) while rectangular grooves are considered in the model. Small
discrepancies between some nominal values and those used in the
calculations are attributed to this shape difference. Grating area
are $2\times2$ $mm^2$. Reflectivity measurements of p-polarized
impinging light were performed at room temperature, in dry air, in
the $1.67$ to $6$ $\mu m$ spectral range, with a BioRad FTS60A
DigiLab Fourier Transform photospectrometer. Spectra were normalized
to the reflectivity of a flat gold surface.
\\Figure 4 displays
reflectivity measurements and calculated spectra realized at
$\theta=10^\circ$. Parameters used in the calculation are:
$h_{<1/1>}= 0.64 \mu m$, $D=2.2$ $\mu m$, $h_{<1/2>}= 0.69 \mu m$,
$S=0.7$ $\mu m$, $L=1.5$ $\mu m$, $h_{<2/3>}= 0.6 \mu m$, $S=0.65
\mu m$, $L=1.3 \mu m$ and $w=0.55 \mu m$ for all samples. Figure 4a
shows the previously measured large cavity mode noted CM as well as
the $n=\pm1$ branches of the surface plasmons noted
SP\cite{nousEPJD}. These SP occurs around the same wave number for
the two other gratings since the period of the three samples have
close values. More interestingly, the splitting of the large CM into
the two predicted resonances for the $<1/2>$ grating and into three
predicted resonances for the $<2/3>$ grating is measured. The three
resonances of $<2/3>$ correspond to the symmetric coupling of the
three cavities ($\rightarrow\rightarrow\rightarrow$), the symmetric
coupling of the two external cavities, anti-symmetric to the central
one whose dipole momentum is twice larger ($\rightarrow
\longleftarrow\rightarrow$) and the anti-symmetric coupling of the
external cavities leading to the extinction of the field in the
central one ($\rightarrow0\leftarrow$).
\begin{figure}
\centering\includegraphics[width=8.5cm]{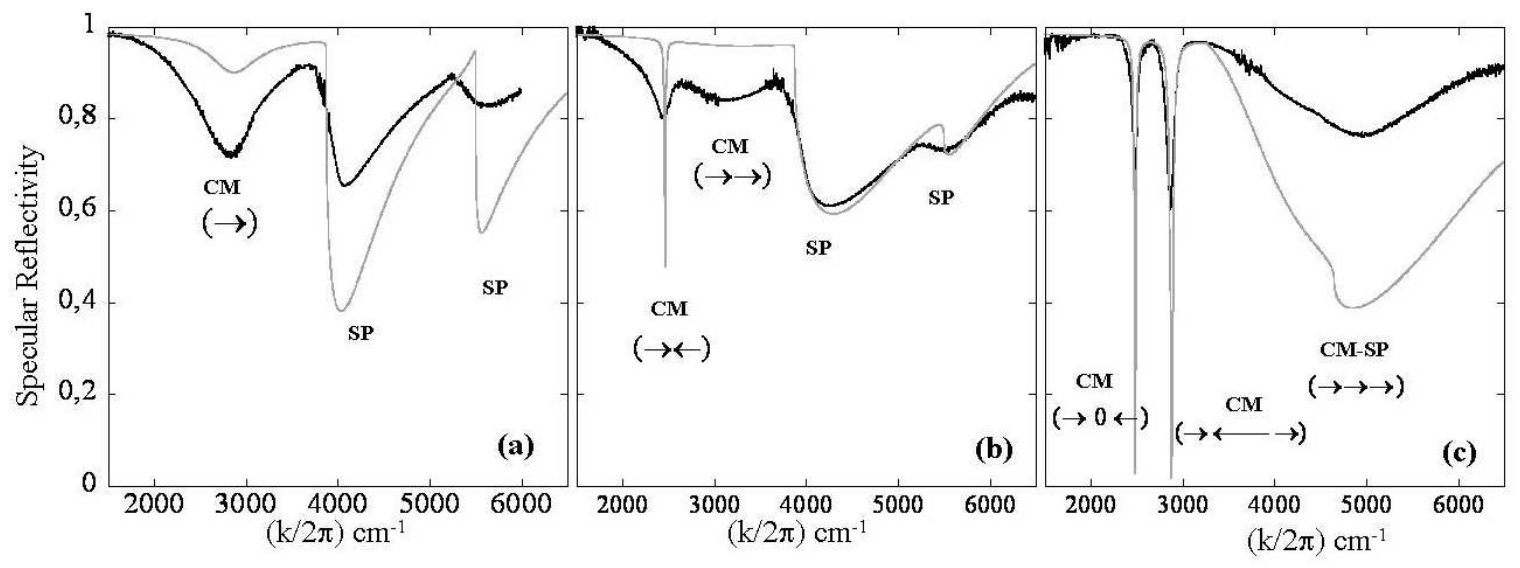}
\caption{Measured (black line) and calculated (light gray line)
specular reflectivity at $\theta=10^\circ$ of the gratings $<1/1>$
(a), $<1/2>$ (b) and $<2/3>$ (c). Cavity resonances are noted CM and
surface plasmon SP. The arrows indicate the direction of the
equivalent dipole momentum at the mouth of each cavity.}
\end{figure}
\\
To highlight the symmetric or anti-symmetric nature of the two sharp
resonances of the $<2/3>$ grating, we have performed reflectivity
measurements at various angles. Increasing $\theta$ breaks the
symmetry of the system and favors the excitation of the
anti-symmetric mode which vanishes at normal incidence (the
symmetric incident field cannot couple to it). Figure 5 evidences
the turnaround of the strength of two resonances as a function of
the incidence angle and demonstrates that peak at smaller wave
number is the anti-symmetric mode ($\rightarrow0\leftarrow$): it is
weakly excited at small incidence angles and grows to become
predominant around $\theta=33^\circ$. Opposite behaviour is observed
for the symmetric dip. Modifying the incident angle may thus be an
easy way to control the near-field distribution.
\begin{figure}
\centering\includegraphics[width=8.5cm]{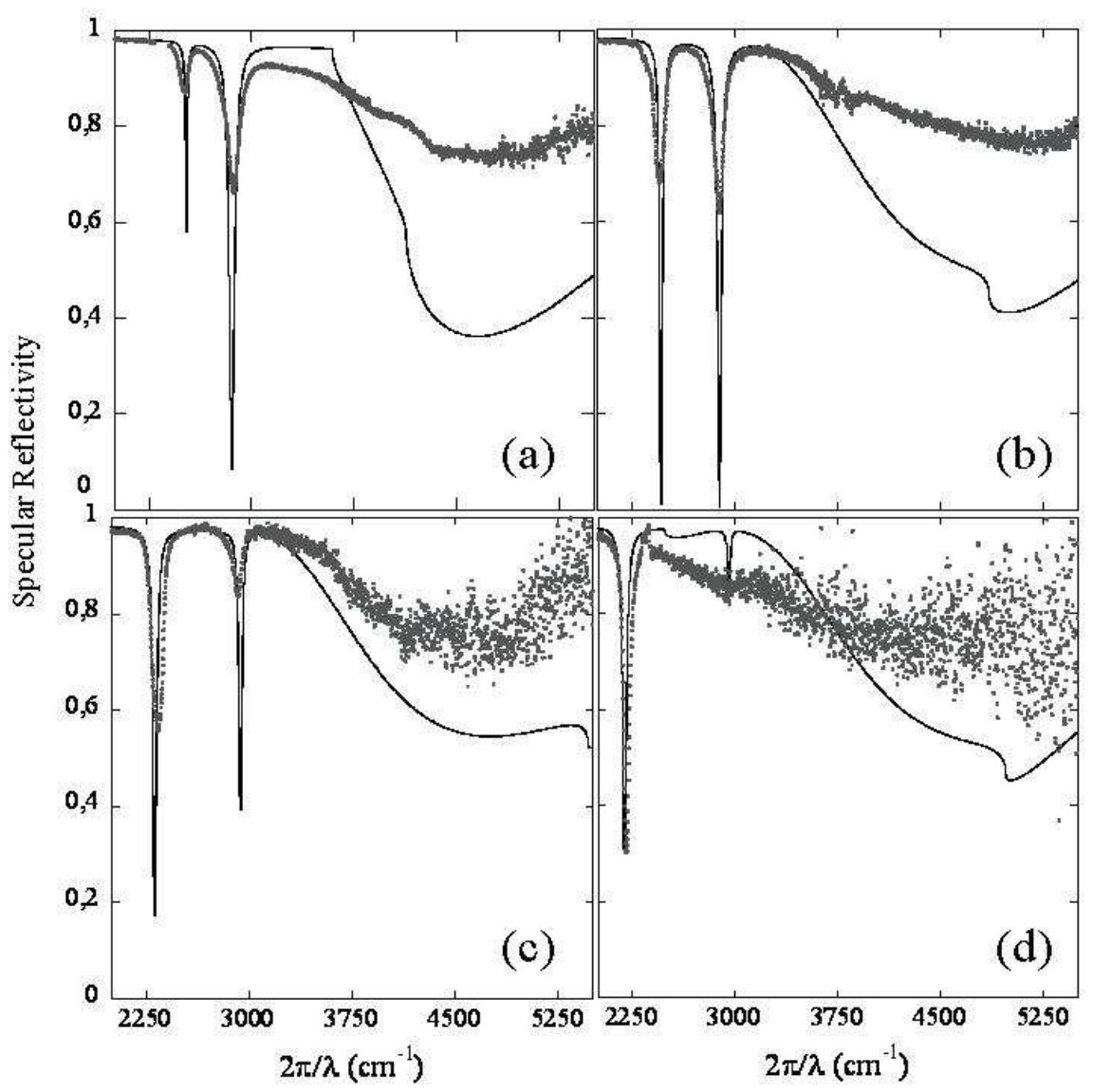}
\caption{Experimental (black line) and calculated (gray line)
specular reflectivity of the $<2/3>$ grating performed at
$\theta=4^\circ$ (a), $\theta=12^\circ$ (b), $\theta=24^\circ$ (c)
and $\theta=33^\circ$ (d). Calculations are made considering four
modes in the cavities.}
\end{figure}\\In conclusion, theoretical study of commensurate
gratings tending to incommensurate structures was used to model and
understand the optical properties of metallic surfaces with complex
topologies. Thanks to these structures we can generate, in a
controlled manner, weakly radiative modes which localize local,
strong and wavelength dependant electromagnetic fields. These
features are of great interest since they are commonly admitted to
occur on disordered and/or SERS surfaces. Finally, experimental
evidence of these modes was given in the
infra-red region for simple commensurate structures.\\

The authors would like to thank Nathalie Bardou for assistance in
the fabrication process.

\begin{references}
%
\bibitem{mayou1} C. Berger, E. Belin, D. Mayou, Ann. Chimi. Mater.
(Paris) \textbf{18}, 485 (1993).
%
\bibitem{mayou2} E. Belin, D. Mayou, Phys. Scr. \textbf{T49}, 356
(1993).
%
\bibitem{mayou3} G. de Laissardiere, D. Nguyen-Manh, D. Mayou, Prog.
Mater. Sci. \textbf{50}, 679 (2005).
%
\bibitem{aubry} S. Aubry and P. Qu\'{e}merais in Low-dimensional electronic
properties of Molybdenum bronzes and oxides, p.293 (Ed. C.
Schlenker, Kluwer Academic publishers 1989).
%
\bibitem{ducastelle} F. Ducastelle, Order and phase stability in
alloys, Cohesion and Structure vol. 3, (Ed. F.R. De Boer, D. G.
Pettifor, North-Holland 1991).
%
\bibitem{quemerais} F. Ducastelle, P. Qu\'{e}merais, Phys. Rev.
Lett. {\bf 78}, 102 (1997).
%
\bibitem{skigin} D. C. Skigin and R. A. Depine, Phys. Rev.
lett. {\bf 95}, 217402 (2005) and references therein.
%
\bibitem{skigin2} D. C. Skigin and R. A. Depine, Phys. Rev.
E {\bf 74}, 046606 (2006).
%
\bibitem{skigin3} D. C. Skigin, A. N. Fantino and S. I. Grosz, J. Opt. A: Pure Appl. Opt. {\bf 5}, 129 (2003).
%
\bibitem{wirgin} A. Wirgin, A.A. Maradudin, Prog. Surf. Sc., {\bf 22}, 1 (1986).
%
\bibitem{nousEPJD} A. Barbara, P. Qu\'emerais, E. Bustarret, T. L\'{o}pez-R\'{\i}os, T. Fournier, Eur. Phys. Jour. {\bf D}, 23, 143 (2003).
%
\bibitem{jap} A. Barbara, P. Qu\'emerais, J. Le Perchec and T. L\'{o}pez-R\'{\i}os,
J. Appl. Phys. {\bf 98}, 033705 (2005).
%
\bibitem{palik} E. D. Palik, \textit{Handbook of Optical Constants of
Solids}, Academic Press.
\bibitem{lopez} T. L\'{o}pez-R\'{\i}os, D. Mendoza, F. J. Garc\'{\i}a-Vidal, J. S\'{a}nchez-dehesa and B. Pannetier, Phys. Rev. Lett. {\bf 81}, 665 (1998).
%
\bibitem{nousPRL} J. Le Perchec, P. Qu\'emerais, A. Barbara and T. L\'{o}pez-R\'{\i}os, Phys. Rev. Lett. {\bf 97}, 036405 (2006).
%
\bibitem{gresillon} S. Ducourtieux et al., Phys. Rev. B \textbf{64},
165403 (2001).
%
\bibitem{sambles} A. P. Hibbins, I. R. Hooper, M. J. Lockyear, J. R. Sambles,
Phys. Rev. Lett. \textbf {96}, 257402 (2006).
%
\bibitem{moskovitz} M. Moskovits, Rev. Mod. Phys. \textbf{57}, 783
(1985).
%
\bibitem{collin} S. Collin, F. Pardo, R. Teissier and J-L. Pelouard  Appl. Phys. Lett \textbf{85},194 (2004).
%
\end{references}
\end{document}